\documentclass[12pt]{article}

\usepackage{graphicx}

\begin{document}

\title{$\tau\rightarrow\omega3\pi\nu$ Decays}
\author{Jun Gao and Bing An Li\\
Department of Physics and Astronomy, University of Kentucky\\
Lexington, KY 40506, USA}

\maketitle

\begin{abstract}
Theoretical study of anomalous decay mode
$\tau\rightarrow\omega\pi\pi\pi\nu$ is presented. Theoretical
value of the branching ratio of
$\tau^-\rightarrow\omega\pi^-\pi^0\pi^0\nu$ agrees well with data.
The branching ratio of
$\tau^{-}\rightarrow\omega\pi^+\pi^-\pi^-\nu_\tau$ is predicted.
It is found that the vertices of $a_{1}\rho\pi$ and
$\omega\rho\pi$ play dominant role in these two decay modes. CVC
is satisfied and there is no adjustable parameter.

\end{abstract}

\newpage
There is rich physics in $\tau$ hadronic decays. Because of the
value of $m_{\tau}$ many light mesons made of u, d, and s quarks,
especially meson resonances, are produced in the decays.
Therefore, $\tau$ mesonic decays provide very unique test ground
of Standard Model and QCD. An effective theory of large $N_{C}$
QCD of mesons has been proposed to study the physics of light
mesons [1]. In this theory the tree diagrams of mesons are at
leading order of large $N_{C}$ expansion and loop diagrams are at
higher orders. This theory has been applied to study many physical
processes of mesons and it has been shown that the theory is
phenomenologically successful [1, 2, 3]. Both vector and
axial-vector currents contribute to $\tau$ decays. The mesonic
vector current is obtained from the Vector Meson Dominance(VMD)
which is a natural result of this theory [1]. The axial-vector
current of mesons is also obtained [3]. PCAC is satisfied[3]. Many
$\tau$ mesonic decay modes have been studied by using this theory
[3]. Theory agrees with data reasonably well.

The decay rate of $\tau^{-}\rightarrow2\pi^-\pi^+3\pi^0\nu_\tau$
has been measured by ALEPH [4] and CLEO [5] and predicted by CVC
[6, 7, 8]. The first measurement of
$\tau^{-}\rightarrow\omega\pi^-\pi^0\pi^0\nu_\tau$ has been
reported by CLEO [5]
\[B(\tau^{-}\rightarrow\omega\pi^-\pi^0\pi^0\nu_\tau)=(1.89^{+0.74}_{-0.67}\pm0.40)\times10^{-4}.\]
There is another decay mode $\tau^-\rightarrow\omega\pi^+\pi^-\pi^-\nu_{\tau}$.
They are very interesting decay modes which are resulted from the
vector current. These decay modes are tests of VMD. $\omega$ meson
is associated with anomaly. Wess-Zumino-Witten anomaly [9] can be
tested by these modes. Since four mesons are produced, many meson
vertices are involved in these processes. These decay modes provide
tests on all kinds of meson theory. In the effective theory of
large $N_{C}$ QCD of mesons [1] all the vertices of these decays
have been derived and all the parameters have been fixed. The
large $N_{C}$ theory of mesons [1] will make definite predictions
on these two decay modes. Therefore, they provide
serious tests on this theory.

In this paper we apply the effective theory of large $N_{C}$ QCD
[1] to study
$\tau^{-}\rightarrow\omega\pi^-\pi^0\pi^0\nu_\tau$ and
$\tau^{-}\rightarrow\omega\pi^+\pi^-\pi^-\nu_\tau$. Only vector
current contributes to both decays, which has been derived as [3]
\begin{equation}
{\cal L}^{V}=\frac{g_{W}}{4}cos\theta_{C}g\{
-{1\over2}(\partial_{\mu}A^{i}_{\nu}-\partial_{\nu}A^{i}_{\mu})(
\partial^{\mu}\rho^{i\nu}-\partial^{\nu}\rho^{i\mu})
+A^{i}_{\mu}j^{i\mu}\},
\end{equation}
where $A^{i}_{\mu}$ is the W boson field,
$j^{i}_{\mu}$ is derived
by the substitution
\begin{equation}
\rho^{i}_{\mu}\rightarrow\frac{g_{W}}{4}gcos\theta_{C}A^{i}_{\mu}
\end{equation}
in the vertices involving $\rho$ meson and g is a universal coupling
constant
which is determined to be 0.39 by fitting $\rho\rightarrow ee^+$.
Eq.(1) is exactly the same expression of VMD given by Sakurai [10].

The diagrams contributing to the decay $\tau\rightarrow \omega
\pi\pi\pi\nu$ are shown in Fig.1(a, b, c, d, e, f). All the
vertices are derived in the chiral limit. The $\omega\rho\pi$
vertex is the Wess-Zumino-Witten anomaly and derived as
\begin{equation}
{\cal L}^{\omega\rho\pi}=-\frac{3}{\pi^{2}g^{2}f_{\pi}}
\varepsilon^{\mu\nu\alpha\beta}\partial_{\mu}
\omega_{\nu}\rho^{i}_{\alpha}\partial_{\beta}\pi^{i}.
\end{equation}
The vertex ${\cal L}^{\omega\rho\pi}$ leads to the Adler-Bell-Jackiw anomaly
of $\pi^0\rightarrow\gamma\gamma$ [1].

Besides the anomalous vertices ${\cal L}^{\omega\rho\pi}$,
there are other three kinds of normal vertices in Fig.1.
The first kind of vertices derived in Ref.[1] are
\begin{eqnarray}
{\cal
L}^{a_{1}\rho\pi}&=&\epsilon_{ijk}\{A(q^{2},p^{2})a^{i}_{\mu}
\rho^{j\mu}\pi^{k}-Ba^{i}_{\mu}\rho^{j}_{\nu}\partial^{\mu\nu}\pi^{k}
+Da^{i}_{\mu}\partial^{\mu}(\rho^{j}_{\nu}
\partial^{\nu}\pi^{k})\}, \\
{\cal L}^{\rho\pi\pi}&=&{2\over g}\epsilon_{ijk}\rho^{i}_{\mu}
\pi^{j}\partial^{\mu}\pi^{k}-{2\over \pi^{2}f^{2}_{\pi}g}
\{(1-{2c\over g})^{2}-4\pi^{2}c^{2}\}\epsilon
_{ijk}\rho^{i}_{\mu}\partial_{\nu}\pi^{j}\partial^{\mu\nu}\pi^{k}
\nonumber \\ &&-{1\over \pi^{2}f^{2}_{\pi}g}\{3(1-{2c\over g})^{2}
+1-{2c\over g}-8\pi^{2}c^{2}\}\epsilon_{ijk}\rho^{i}_{\mu}\pi_{j}
\partial^{2}\partial_{\mu}\pi_{k},\\
{\cal L}^{\rho\rho\rho}&=&-{2\over g}\epsilon_{ijk}\partial_{\mu}
\rho^{i}_{\nu}\rho^{j\mu}\rho^{k\nu},
\end{eqnarray}
where
\[c=\frac{f^2_{\pi}}{2gm^2_{\rho}},\]
\[F^2=(1-{2c\over g})^{-1}f^2_{\pi},\]
\begin{eqnarray}
A(q^{2},p^{2})&=&{2\over f_{\pi}}f_{a}\{{F^{2}\over g^{2}}
+p^{2}[{2c\over g}+{3\over4 \pi^{2}g^{2}}(1-{2c\over g})]\nonumber
\\ &&+q^{2}[{1\over 2\pi^{2}g^{2}}- {2c\over
g}-{3\over4\pi^{2}g^{2}}(1-{2c\over g})]\},\\
f_{a}&=&(1-{1\over2\pi^{2}g^{2}})^{-{1\over2}},\\ B&=&-{2\over
f_{\pi}}f_{a}{1\over2\pi^{2}g^{2}}(1-{2c\over g}),\\
D&=&-{2\over f_{\pi}}f_{a}\{{2c\over g}+{3\over2\pi^{2}g^{2}
}(1-{2c\over g})\}
\end{eqnarray}
with q being the momentum of $a_1$ meson and p the momentum of
$\rho $ meson.

The vertices of ${\cal L}^{a_1\rho\pi}$
and ${\cal L}^{\rho\pi\pi}$ have been tested by the widths of $a_1$ and $\rho$
, pion form factors, and other physical processes [1, 2, 3]. Theory agrees
well
with data.

The second kind of normal vertices are contact interactions between two meson
fields
\begin{eqnarray}
{\cal L}^{\pi\pi}&=&\frac{1}{4\pi^{2}f^{2}_{\pi}} (1-{2c\over
g})^{2}\partial_{\mu\nu}\pi^{i}
\partial^{\mu\nu}\pi^{i},\\
{\cal L}^{\pi a}&=&\frac{f_{a}}{2\pi^{2}gf_{\pi}} (1-{2c\over
g})\partial_{\mu\nu} \pi^{i}\partial^{\mu}a^{i\nu},\\ {\cal
L}^{aa}&=&\frac{f^{2}_{a}}{4\pi^{2}g^{2}}(\partial_{\mu}a^{i\mu})^{2}.
\end{eqnarray}
By inserting these vertices into Fig.1(b, c), related diagrams are
obtained.

The third kind of vertex is the direct interaction between $\rho
\rho \pi \pi $
\begin{eqnarray}
{\cal L}^{\pi\pi\rho\rho}&=&{4\over f_{\pi}}\epsilon_{ijk}
\epsilon_{i
j'k'}\{\frac{F^{2}}{2g^{2}}\rho^{j}_{\mu}\rho^{j'}_{\mu}
\pi_{k}\pi_{k'}+[-{2c^{2}\over
g^{2}}+\frac{3}{4\pi^{2}g^{2}}(1-{2c\over g})^{2}]
\rho^{j}_{\mu}\rho^{j'}_{\nu}\partial^{\mu}\pi_{k}\partial^{\nu}\pi_{k'}
\nonumber \\ &&+[-{2c^{2}\over
g^{2}}+\frac{1}{4\pi^{2}g^{2}}(1-{2c\over g})^{2}]
\rho^{i}_{\nu}\rho^{j'\nu}\partial_{\mu}\pi_{k}\partial^{\mu}\pi_{k'}
-{2c^{2}\over
g^{2}}\rho^{j}_{\mu}\rho^{k'}_{\nu}\partial^{\nu}\pi_{k}
\partial^{\mu}\pi_{j'}\nonumber \\
&&-{3\over2\pi^{2}g^{2}}(1-{2c\over
g})(\rho^{j}_{\mu}\pi_{k}\partial_{\nu}
\pi_{j'}-\rho^{j}_{\nu}\pi_{k}\partial_{\mu}\pi_{j'})\partial^{\nu}
\rho^{k'\mu}-{1\over2\pi^{2}g^{2}}(1-{2c\over
g})\rho^{j}_{\mu}\pi_{k}
\rho^{j'}_{\nu}\partial^{\mu\nu}\pi_{k'}\nonumber \\
&&+{1\over4\pi^{2}g^{2}}[\partial_{\nu}(\rho^{j}_{\mu}\pi_{k})
\partial^{\nu}(\rho^{j}_{\mu}\pi_{k'})+2(1-{2c\over g})\rho^{j}_{\mu}\pi_{k}
\rho^{j'}_{\nu}\partial^{\mu\nu}\pi_{k'}]\}.
\end{eqnarray}
This is the vertex of Fig.1(a).

Due to the structure of the vertex (1) the amplitudes of the
diagrams Fig.1(e, f) satisfy the CVC automatically. However, 
for the diagrams(Fig.1(a, b, c, d)) we have
to put all the three kinds of vertices(4-6, 11-14) together to
make CVC be satisfied in the chiral limit. These vertices have
been exploited to calculate the branching ratios of
$\tau\rightarrow\rho\pi\pi\nu$ [11]. Theoretical results are in
agreement with the data. It is necessary to emphasize that all the
parameters of this study have been fixed in previous studies.
There is no adjustable parameter in this investigation.

In diagram Fig.1(e) there are two anomalous vertices ${\cal
L}^{\omega\rho\pi}$. The study [1] shows that the strength of this
vertex is weaker than normal vertices. This is the reason why
$\omega$ meson has narrower decay width than $\rho$ meson. The
calculation shows that the contribution of diagram Fig.1(e) is
negligible. In the diagram Fig.1(f) there are $\rho$ resonance,
$\omega$ and $\pi$ mesons in the final state. Because the phase
space of this process is too small the contribution of this
diagram is also negligible.

Now we calculate the branching ratios of
$\tau\rightarrow\omega\pi\pi\pi\nu$. It is very lengthy. The
amplitudes of the decays $\tau\rightarrow\omega\pi\pi\pi\nu$ are
obtained from all the three kinds of vertices(4-6, 11-14) and the
vertex ${\cal L}^{\omega\rho\pi}$ (3). In the chiral limit the
matrix elements of the vector current of
$\tau^{-}\rightarrow\omega\pi^-\pi^0\pi^0\nu_\tau$ and
$\tau^{-}\rightarrow\omega\pi^+\pi^-\pi^-\nu_\tau$ have been
derived as
\begin{eqnarray}
&&<\omega (p)\pi ^0(p_1)\pi ^0(p_2)\pi ^{-}(p_3)\mid j ^{\mu-}\mid
0> \nonumber \\ &&=\frac 1{\sqrt{16E\omega _1\omega _2\omega _3}}
(g^{\mu\lambda}-\frac{q^{\mu}q^{\lambda}}{q^2}) \frac 3{\pi
^2gf_\pi }
\frac{-m^2_{\rho}+i\sqrt{q^2}\Gamma_{\rho}(q^2)}{q^2-m^2_{\rho}+i\sqrt{q^2}
\Gamma_{\rho}(q^2)}f^{00}_{\lambda},\\ &&f^{00}_{\lambda}=
\epsilon _{\nu'}p_{\mu'} \varepsilon ^{\mu' \nu'\alpha \beta}
\nonumber \\ &&\{\frac{p_{3\beta }}{(p+p_{3 })^2-m_\rho
^2+i\sqrt{(p+p_{3 })^2}\Gamma _\rho ((p+p_{3 })^2)}(f_{\alpha
\lambda }^{(12)}(p+p_{3 })+f_{\alpha \lambda }^{(21)}(p+p_{3 }))
\nonumber
\\ &&-\frac{p_{1\beta }}{(p+p_{1 })^2-m_\rho ^2+i\sqrt{(p+p_{1
})^2}\Gamma _\rho ((p+p_{1 })^2)}f_{\alpha \lambda }^{(23)}(p+p_{1
}) \nonumber \\ &&-\frac{p_{2\beta }}{(p+p_{2 })^2-m_\rho
^2+i\sqrt{(p+p_{2 })^2}\Gamma _\rho ((p+p_{2 })^2)}f_{\alpha
\lambda }^{(13)}(p+p_{2 })\};  \\ \nonumber \\&&<\omega (p)\pi
^{+}(p_2)\pi ^{-}(p_1)\pi ^{-}(p_3)\mid j ^{\mu-}\mid 0> \nonumber
\\ &&=\frac 1{\sqrt{16E\omega _1\omega _2\omega _3}}
(g^{\mu\lambda}-\frac{q^{\mu}q^{\lambda}}{q^2}) \frac 3{\pi
^2gf_\pi }
\frac{-m^2_{\rho}+i\sqrt{q^2}\Gamma_{\rho}(q^2)}{q^2-m^2_{\rho}+i\sqrt{q^2}
\Gamma_{\rho}(q^2)}f^{--}_{\lambda},
\\ &&f^{--}_{\lambda}=
\epsilon _{\nu'}p_{\mu'} \varepsilon ^{\mu' \nu'\alpha \beta}
 \nonumber
\\ &&\{\frac{p_{3\beta }}{(p+p_{3 })^2-m_\rho
^2+i\sqrt{(p+p_{3 })^2}\Gamma _\rho ((p+p_{3
})^2)}f_{\alpha \lambda }^{(12)}(p+p_{3 }) \nonumber
\\&&+\frac{p_{1\beta }}{(p+p_{1 })^2-m_\rho
^2+i\sqrt{(p+p_{1})^2}\Gamma _\rho ((p+p_{1 })^2)}f_{\alpha
\lambda }^{(32)}(p+p_{1 }) \nonumber \\ &&-\frac{p_{2\beta
}}{(p+p_{2 })^2-m_\rho ^2+i\sqrt{(p+p_{2 })^2}\Gamma _\rho
((p+p_{2 })^2)}(f_{\alpha \lambda }^{(13)}(p+p_{2 })+f_{\alpha
\lambda }^{(31)}(p+p_{2 }))\},
\end{eqnarray}
where
\begin{eqnarray}
q&=&p+p_1+p_2+p_3, \\ f_{\alpha \lambda }^{(12)}(p_\rho
)&=&g_{\alpha \lambda }f(p_\rho )+p_{1\alpha }[f_{11}(p_\rho
)p_{1\lambda }+f_{21}(p_\rho )p_{2\lambda }] \nonumber \\
&&+p_{2\alpha }[f_{12}(p_\rho )p_{1\lambda }+f_{22}(p_\rho
)p_{2\lambda }]
\end{eqnarray}
with $p_\rho $ being the momentum of the $\rho $ meson, $p_i$ (i =
1, 2, 3) the momentum of pion, and p the momentum of $\omega$.
$f_{\alpha \lambda }^{(21)}$ is obtained from $f_{\alpha \lambda
}^{(12)}$ by exchanging $p_1\leftrightarrow p_2$; $f_{\alpha
\lambda }^{(23)}$ is obtained from $f_{\alpha \lambda }^{(21)}$ by
replacing $p_1$ with $p_3$; $f_{\alpha \lambda }^{(13)}$ is
obtained from $f_{\alpha \lambda }^{(12)}$ by replacing $p_2$ with
$p_3$; $f_{\alpha \lambda }^{(32)}$ is obtained from $f_{\alpha
\lambda }^{(23)}$ by exchanging $p_2\leftrightarrow p_3$;
$f_{\alpha \lambda }^{(31)}$ is obtained from $f_{\alpha \lambda
}^{(13)}$ by exchanging $p_1\leftrightarrow p_3$. $\Gamma _\rho $
is the decay width of $\rho $ meson of momentum q
\begin{equation}
\Gamma_{\rho}(q^{2})=\frac{\sqrt{q^2}}{12\pi g^2}\{1+
\frac{q^{2}}{2\pi^{2}f^{2}_{\pi}}[(1-{2c\over g})^{2}-
4\pi^{2}c^{2}]\}^{2}(1-{4m^{2}_{\pi}\over q^{2}})^{3\over2}.
\end{equation}
Eqs.(15, 17) show that the CVC, indeed, is satisfied in the chiral limit.
It is interesting to notice that the resonance factor
\[
\frac{-m^2_{\rho}+i\sqrt{q^2}\Gamma_{\rho}(q^2)}{q^2-m^2_{\rho}+i\sqrt{q^2}
\Gamma_{\rho}(q^2)}\] in Eqs.(15, 17) is obtained from the
combination of the two terms in Eq.(1). These two terms are shown
in two diagrams of Fig.1(a, b, c, d, e, f).

The contributions of $a_1$ meson (Fig.1(b)) to the functions f and
$f_{ij}$ (i, j = 1, 2) are given below. The contributions from
other diagrams are shown in the appendix.
\begin{eqnarray}
BW(k_1^2)&=&\frac 1{k_1^2-m_a^2+i\sqrt{k_1^2}\Gamma _a(k_1^2)},
\nonumber \\f(p_\rho )&=&BW(k_1^2)A(q^2,k_1^2)A(p_\rho ^2,k_1^2),
\nonumber
\\f_{11}(p_\rho )&=&BW(k_1^2)A(p_\rho ^2,k_1^2)B, \nonumber \\
f_{12}(p_\rho )&=&BW(k_1^2)\{[A(p_\rho
^2,k_1^2)+A(q^2,k_1^2)]D+(k_1\cdot p_2-k_1\cdot p_1)BD+p_1\cdot
p_2B^2-k_1^2D^2\} \nonumber \\ &&-BW(k_1^2)\frac
1{m_a^2}\{-A(q^2,k_1^2)+k_1\cdot p_1B+k_1^2D\}\{A(p_\rho
^2,k_1^2)+k_1\cdot p_2B-k_1^2D\}, \nonumber
\\f_{22}(p_\rho )&=&BW(k_1^2)A(q^2,k_1^2)B,
\end{eqnarray}
where $k_i=q-p_i$ (i = 1, 2, 3) and
\begin{equation}
m^2_a=(\frac{F^2}{g^2}+m^2_{\rho})/(1-{1\over2\pi^2 g^2}).
\end{equation}
The decay width
of $a_{1}$ meson is derived as
\begin{eqnarray}
\Gamma_{a}(k^{2})&=&{k_{a}\over 12\pi}{1\over k^{2}}
\{(3+{k^{2}_{a}\over m^{2}_{\rho}})A^{2}(m^{2}_{\rho},k^{2})
\nonumber \\
&&-A(m^{2}_{\rho},k^{2})B(k^{2}+m^{2}_{\rho}){k^{2}_{a} \over
m^{2}_{\rho}}+{k^{2}\over m^{2}_{\rho}}k^{4}_{a}B^{2}\}, \nonumber
\\ k^{2}_{a}&=&{1\over 4k^{2}}(k^{2}+m^{2}_{\rho}-m^{2}_{\pi})^2
-m^{2}_{\rho}.
\end{eqnarray}

The decay rate of $\tau\rightarrow\omega\pi\pi\pi\nu$ is derived from Eqs.(15-18)
\begin{equation}
\frac{d\Gamma^{ab}}{dq^2}={1\over128}{G^2\over(2\pi)^8}cos^2\theta_{c}{1\over
q^2} {1\over
m^3_{\tau}}(m^2_{\tau}-q^2)^2(m^2_{\tau}+2q^2)(\frac3{\pi^2gf_{\pi}})^2
\frac{m^4_{\rho}+q^2\Gamma^2_{\rho}(q^2)}{(q^2-m^2_{\rho})^2+q^2\Gamma^2
_{\rho}(q^2)}F^{ab}(q^2),
\end{equation}
where
\[ab=00\;\; or\;\; --,\]
\begin{equation}
F^{ab}(q^2)={1\over3}
(g^{\lambda\lambda'}-\frac{q^{\lambda}q^{\lambda'}}{q^2}){1\over4(2\pi)^2}
\int\frac{d^3p_1d^3 p_2 d^3 p_3 d^3 p}{E\omega_1\omega_2\omega_3}
\delta(q-p_1-p_2-p_3-p) f^{ab}_{\lambda}f^{*ab}_{\lambda'}.
\end{equation}

The branching ratios of the two decay channels are calculated to be
\begin{eqnarray}
B(\tau ^{-}\rightarrow \omega\pi ^{-}\pi ^0\pi ^0\nu _\tau
)&=&2.16\times 10^{-4}, \\ B(\tau ^{-}\rightarrow \omega\pi
^{+}\pi ^{-}\pi ^{-}\nu _\tau )&=&2.18\times 10^{-4}.
\end{eqnarray}
Theoretical branching ratio of
$\tau^-\rightarrow\omega\pi^-\pi^0\pi^0\nu_\tau$ is consistent
with the data [5]. The theory predicts that the branching ratio of
$\tau^-\rightarrow\omega\pi^+\pi^-\pi^-\nu_\tau$ is about the same
as that of $\tau^-\rightarrow\omega\pi^-\pi^0\pi^0\nu_\tau$.

As shown in Fig.1, there are many subprocesses in the decays.
However, the calculation shows that the $a_1$ meson (Fig.1(b))
dominates the two decay channels. If only the subprocess which is
obtained from ${\cal L}^{a_1\rho\pi}$ is kept in the matrix
elements (15, 17), we obtain
\begin{eqnarray}
B(\tau ^{-}\rightarrow \omega\pi ^{-}\pi ^0\pi ^0\nu _\tau
)&=&1.86\times 10^{-4}, \\ B(\tau ^{-}\rightarrow \omega\pi
^{-}\pi ^{-}\pi ^{+}\nu _\tau )&=&1.87\times 10^{-4}.
\end{eqnarray}
$86\%$ of the decay rate comes from ${\cal L}^{a_1\rho\pi}$.
It is necessary to point out that 
in Fig.1(b) there are terms which violate CVC. However, these terms are cancelled by the corresponding terms of other diagrams.
The results(29, 30) are obtained
after this cancellation.
It is interesting to notice that $a_1$ meson is associated with the $a_1$
dominance in the axial-vector current [3].

The distribution functions of the two decay modes, $\frac{d\Gamma
}{dq^2}$, are calculated and shown in Fig.2 and Fig.3. There is a
peak in each distribution, which originates in the combination of
the $a_{1}$ resonance and the kinematics of the decays.

To conclude, two decay modes of
$\tau\rightarrow\omega\pi\pi\pi\nu$ have been studied by an
effective large $N_{C}$ QCD of mesons. CVC is satisfied in the
chiral limit. Theoretical branching ratio of
$\tau^{-}\rightarrow\omega\pi^-\pi^0\pi^0\nu_\tau$ agrees with the
data. The theory predicts that the branching ratio of
$\tau^-\rightarrow\omega\pi^+\pi^-\pi^-\nu_\tau$ is about the same
as that of $\tau^-\rightarrow\pi^-\pi^0\pi^0\nu_\tau$. In this
study there is no adjustable parameter.
\newpage
\leftline{\large Appendix}
\begin{enumerate}
\item diagrams involving the vertices (11-13)
\begin{eqnarray}
f_{12}(p_\rho )&=&\frac 1{2\pi ^2g^2}f_a^2\frac
1{m_a^4}\{-A(q^2,k_1^2)+k_1\cdot p_1B+k_1^2D\} \nonumber \\
&&\times \{A(p_\rho ^2,k_1^2)+k_1\cdot p_2B-k_1^2D\} \nonumber \\
&&+\frac 1{2\pi ^2gf_\pi }f_a\frac
1{m_a^2}(1-\frac{2c}g)\{-A(q^2,k_1^2)+k_1\cdot p_1B+k_1^2D\}
\nonumber \\ &&\times\{2F_1(-k_1\cdot p_2)-k_1^2F_2\} \nonumber \\
&&-\frac 1{2\pi ^2gf_\pi }f_a(1-\frac{2c}g)\frac
1{m_a^2}\{A(p_\rho ^2,k_1^2)+k_1\cdot p_2B-k_1^2D\} \nonumber \\
&&\times\{2F_1(k_1\cdot p_1)-k_1^2F_2\} \nonumber \\ &&-\frac
1{2\pi ^2f_\pi ^2}(1-\frac{2c}g)^2\{2F_1(k_1\cdot p_1)-k_1^2F_2\}
\nonumber \\ &&\times\{2F_1(-k_1\cdot p_2)-k_1^2F_2\},
\end{eqnarray}
where
\begin{eqnarray}
F_1(q_1\cdot q_2)&=&\frac 2g\{1+\frac 1{\pi ^2f_\pi ^2}q_1\cdot
q_2[(1-\frac{2c}g)^2-4\pi ^2c^2]\}, \nonumber \\ F_2&=&\frac
4{f_\pi ^2}[\frac{2c^2}g-\frac 3{4\pi ^2g}(1-\frac{2c}g)^2-\frac
1{4\pi ^2g}(1-\frac{2c}g)].
\end{eqnarray}
\item diagrams (Fig.1(c))
\begin{eqnarray}
f_{12}(p_\rho )&=&-\frac 1{k_1^2}\{-4F_1(k_1\cdot
p_1)F_1(-k_1\cdot p_2) \nonumber \\ &&+2k_1^2F_2F_1(-k_1\cdot
p_2)+2k_1^2F_2F_1(k_1\cdot p_1)-k_1^4F_2^2\}.
\end{eqnarray}
\item diagrams (Fig.1(a))
\begin{eqnarray}
f(p_\rho )&=&\frac 4{f_\pi ^2}\{\frac{F^2}{g^2}-2p_1\cdot
p_2[-\frac{2c^2}{g^2}+\frac 1{4\pi ^2g^2}(1-\frac{2c}g)^2]
\nonumber \\ &&-\frac 3{2\pi ^2g^2}(1-\frac{2c}g)(p_2\cdot p_\rho
-q\cdot p_1)+\frac{k_1^2}{2\pi ^2g^2}\}, \nonumber \\
f_{11}(p_\rho )&=&-{2\over\pi^{2}g^{2}f^{2}_{\pi}}(1-{2c\over
g}),\nonumber
\\ f_{12}(p_\rho )&=&-{8\over f^{2}_{\pi}}\{-{2c^{2}\over
g^{2}}+{3\over4\pi^{2}g^{2}} (1-{2c\over
g})^{2}+{3\over2\pi^{2}g^{2}}(1-{2c\over g})\},\nonumber \\
f_{21}(p_\rho )&=&{8\over f^{2}_{\pi}}\{-{4c^{2}\over
g^{2}}+{3\over4\pi^{2}g^{2}} (1-{2c\over g})\},\nonumber \\
f_{22}(p_\rho )&=&f_{11}(p_\rho ).
\end{eqnarray}
\item diagrams (Fig.1(d))
\begin{eqnarray}
BW&=&\frac 1{(p_1+p_2)^2-m_\rho ^2+i\sqrt{(p_1+p_2)^2}\Gamma _\rho
((p_1+p_2)^2)}, \nonumber \\ f(p_\rho )&=&\frac 8{g^2}BWq\cdot
(p_2-p_1)\{1+\frac{p_1\cdot p_2}{\pi ^2f_\pi
^2}[(1-\frac{2c}g)^2-4\pi ^2c^2]\}, \nonumber \\ f_{12}(p_\rho
)&=&\frac{16}{g^2}BW\{1+\frac{p_1\cdot p_2}{\pi ^2f_\pi
^2}[(1-\frac{2c}g)^2-4\pi ^2c^2]\}, \nonumber \\ f_{21}(p_\rho
)&=&-f_{12}(p_\rho ).
\end{eqnarray}
\end{enumerate}

\leftline{The authors wish to thank M.Barnett. The study is
supported by DOE grant No.DE-91ER75661.}

\pagebreak

\pagebreak
\begin{flushleft}
{\bf Figure Captions}
\end{flushleft}
{\bf FIG. 1.} Feynman Diagrams of $\tau \rightarrow \omega \pi \pi
\pi \nu $.
\\ {\bf FIG. 2.} Distribution function $\frac{d\Gamma }{dq^2}$ for
$\tau ^{-}\rightarrow \omega \pi ^{+}\pi ^{-}\pi ^{-}\nu _\tau $.
\\{\bf FIG. 3.} Distribution function $\frac{d\Gamma }{dq^2}$ for
$\tau ^{-}\rightarrow \omega \pi ^{-}\pi ^0\pi ^0\nu _\tau $.

\begin{figure}
\begin{center}
\includegraphics[width=7in, height=7in]{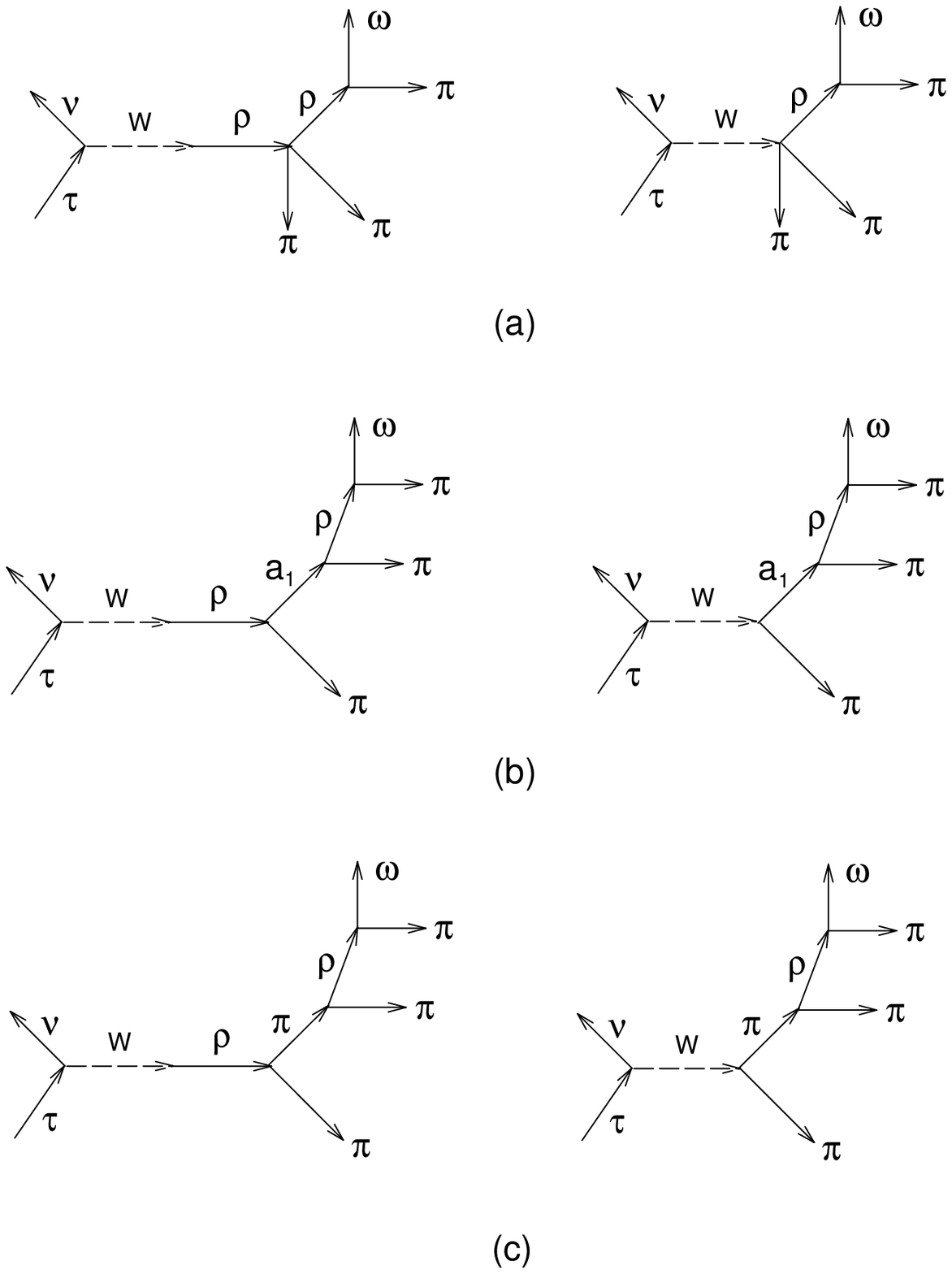}
FIG. 1.
\end{center}
\end{figure}

\begin{figure}
\begin{center}
\includegraphics[width=7in, height=7in]{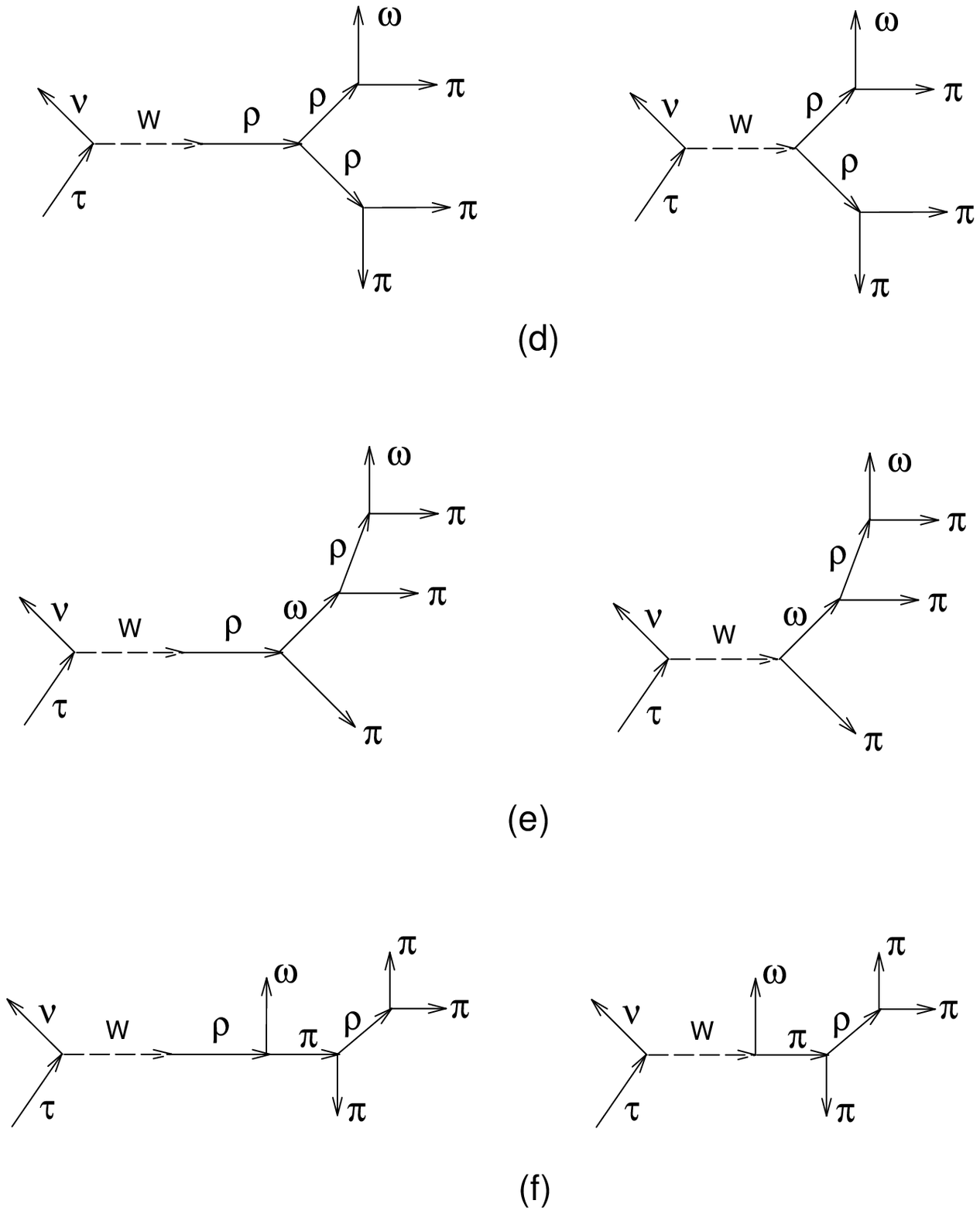}
FIG. 1 cont'.
\end{center}
\end{figure}

\begin{figure}
\begin{center}
\includegraphics[width=7in, height=7in]{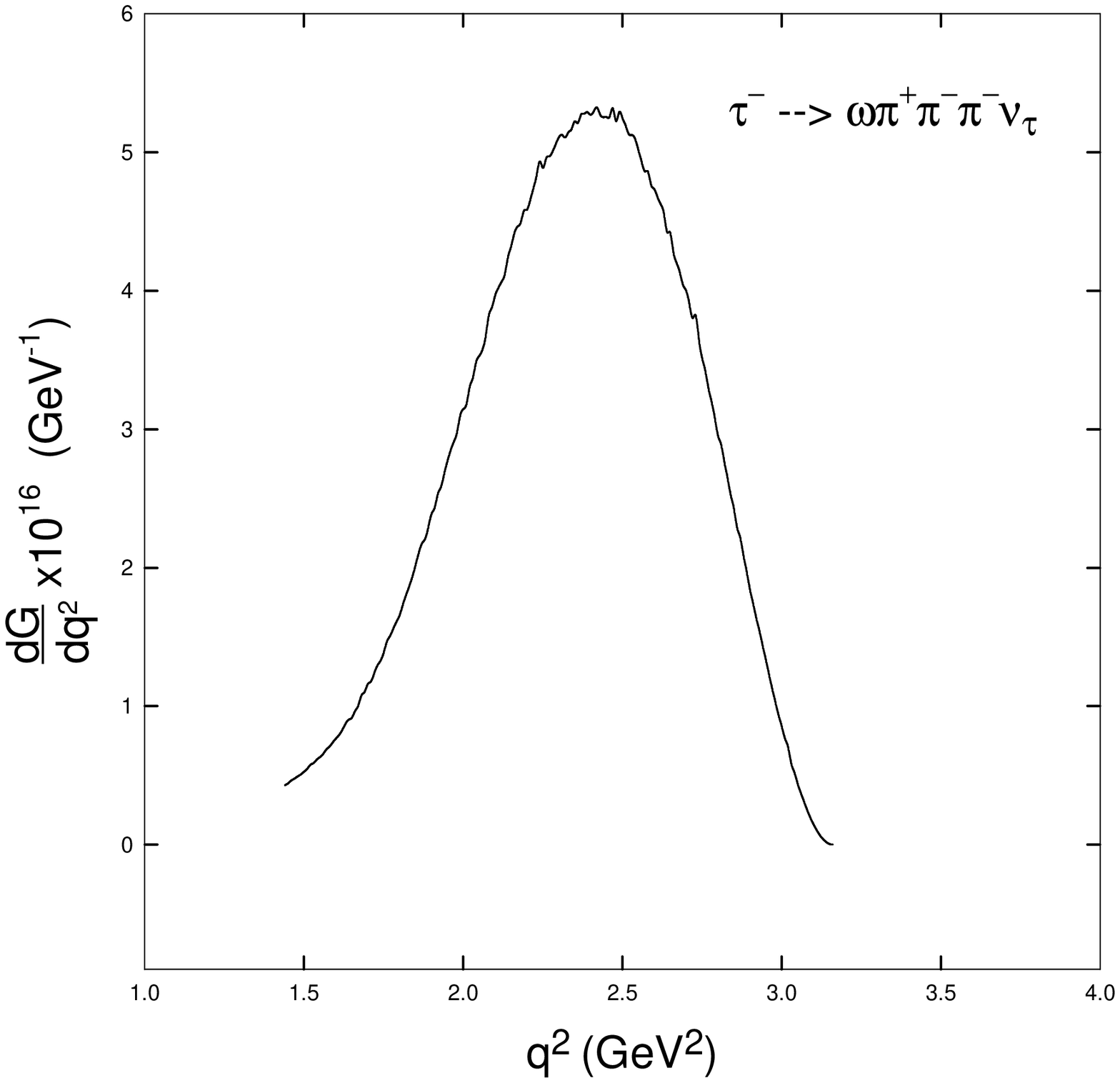}
FIG. 2.
\end{center}
\end{figure}

\begin{figure}
\begin{center}
\includegraphics[width=7in, height=7in]{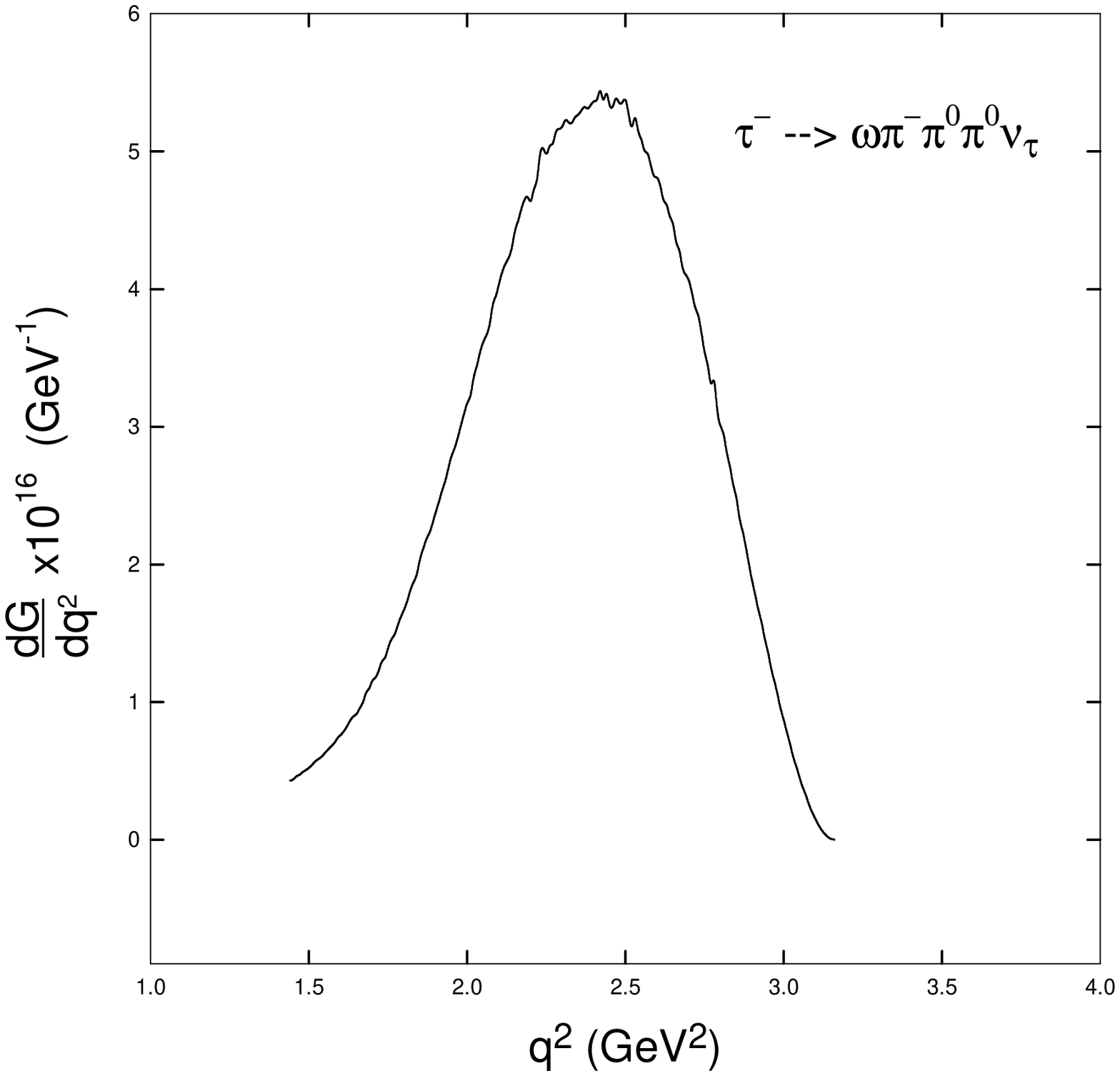}
FIG. 3.
\end{center}
\end{figure}

\end{document}